# Bipolar Almost Equiangular Tight Frames For NOMA Systems

by

Lev Gurevich

# Abstract

Non-Orthogonal Multiple Access (NOMA) is a concept which is gaining a big popularity in multiuser networks. It's due to its advantages in sense of total network throughput. It becomes especially significant in large networks such as Internet of Things (IoT) networks or 5G networks. One of the known NOMA techniques is DS-CDMA NOMA, which make use of non-orthogonal coding schemes to optimize capacity at multiuser networks. Equiangular Tight Frames (ETF) are known to be an optimal sequences' sets (in sense of capacity) for this technique. Unfortunately, ETFs are limited to very specific pairs of users' number and sequence lengths which put undesirable constraints on practical systems. In this paper our goal is to break those constraints by proposing alternative family of non-orthogonal sequences which on the one hand, possess similar properties to those of ETFs (for that reason we'll denote them as Almost ETFs) and on the other hand, doesn't have those limitation on users' number and sequence length. We're basing our approach by starting with known technique of building standard ETFs, and extending it by slight modifications to technique of building AETFs. In this paper we'll concentrate on bipolar (+/-1 valued) and relatively short (up to length of 100) sequences, since we're interested in sequences which will be of practical value in real systems.

# Contents

# 1. Introduction

A modern world of communication is characterized by rapidly growing data traffic and demand for multiuser communication with a large number of devices. The concept of IoT (Internet of Things) is a representative example of such a trend. The IoT refers to network of devices with sensors, processing ability, transceivers which communicate with each other and exchange various data. Those networks are usually characterized by a large number of devices (end users), which are often autonomous, and the amount of traffic in those networks may be large. It presents a new challenges in designing of such a networks – the essence of those challenges is how to design a network between huge amount of users and huge traffic between them such that performance will be optimized (usually in the sense of capacity).

A classical approach to tackle a multiuser communication networks is called Orthogonal Multiple Access (OMA) which allocate for each user an resource (frequency band, time slot, spreading sequence etc.) which is orthogonal to other users' resources – in such a way an interuser interference is prevented (in the ideal case). The main drawback of that approach is that number of orthogonal resources is very limited (and as a consequence, given desired throughput per user, number of users in such a network is limited), because orthogonality constraint between resources presents a limit on possible spectral efficiency per user that can be achieved in such a network. This drawback is very painful for systems with large number of users and large traffic requirements like IoT. In additional, taking into account that in standard networks at any given time instance only small percentage of users are active, allocation of an orthogonal resource to each user (which mostly will be unused) is basically waste of resources. One of the possible alternatives is using Non-Orthogonal Multiple Access (NOMA) – in this approach an orthogonality constraint is replaced by a much looser one – the multiuser interference is allowed but controlled and properly tuned in accordance with desired performance and users' activity. This allows to significantly increase the number of allocated resources and as a consequence may support much larger number of users.

In our paper we concentrate on NOMA where each user is allocated a non-orthogonal sequence. For a given number of users N and given sequence length M (M<N), there a lot of possibilities to build non orthogonal sequences, but the challenge here is to build such a sequences that will minimize interuser interference (given that only K users are active at each time instance). Much of research was made on that topic and it was found that indeed there exist specific family of non-orthogonal sequences that possess this property of optimal interference. A frame (MxN matrix) which consists of vectors of that optimal family is known as Equiangular Tight Frame (ETF) (see section 3.1). A good overview of ETFs is given in [1].

In previous work [2] the performance of different types of complex valued frames was examined (in terms of capacity).

The goal of this paper is to present and analyze practical frames which give near optimal performance. The "practical" aspect embodied in a fact that in practical systems we deal with large but finite number of users – hence we'll investigate cases up to 100 users. In additional "practicality" referred to the fact that we'll deal with bipolar frames (+/-1 valued frames) which may be beneficial for practical transceivers as it simplifies an implementation as compared to the case with complex (or even real) values ETFs. "Near-Optimality" is referred to the fact that although we know that ETFs are the optimal frames, our method of ETFs generation(using Difference Sets) is limited to very specific values of M and N – hence in order to generate frames for any desired (N,M) pair we'll have to present the concept of Almost ETF(AETF), which will be described in details in next sections, but in essence AETF are frames with near ETFs properties but with much more flexible construction. The price is, of course, - performance degradation, but as it will be shown, presented degradation won't have a drastic impact on the performance.

## 2. Problem Definition

We're dealing with Non Orthogonal Multiple Access(NOMA) communication network. We have N users where each user is allocated M resources(M<N) and at each time instance only K users are active(partially NOMA). The whole setup can be described as follows:

$$\underline{y} = c\tilde{F}\underline{\tilde{x}} + \underline{n} \quad (2.1)$$

Where:

- $\underline{x} \sim CN(0, I_N)$ is an N dimensional vector with N signals from N users.
- $\underline{\tilde{x}}$ is a K dimensional vector which represents K out of N active users.
- $F$ is $M \times N$ frame which allocates for each of the users M's length sequence – namely $i^{th}$ user is allocated $\underline{f_i}$ sequence.

$$F = (\underline{f_1}, \underline{f_2}, \dots, \underline{f_N})$$

- $\tilde{F}$ is $M \times K$ subframe where K is a number of active users in that specific time instance.
- $\underline{n} \sim CN(0, I_M)$ is a noise vector of length M.
- $c$ is a normalization factor that ensures desired SNR value. Namely:

$$E\left[\left\|c\underline{f_i}x_i\right\|^2\right] = SNR, \forall i = 1, \dots, N$$

Partially NOMA is a system where at each time instance only K out of N users are transmitting. We're modeling it in our simulation by randomly choosing K out of N users at each iteration.

It'll be convenient to define few denotations that will be frequently used through the paper.

- $\gamma = \frac{M}{N}$ - resources to total users ratio. ($0 < \gamma < 1$)
- $\beta = \frac{K}{M}$ – active users to resources ratio. ($\beta > 0$)
- $p = \frac{K}{N}$ – part of active users out of total number of users. ($0 < p < 1$)
- Relation between the three is $p = \beta\gamma$

# 3. Background
## 3.1. Frames Theory

Only the basic concepts are presented here – for more details reader is referred to [1].

A matrix $F \in \mathbb{C}^{M \times N}$ with $M \leq N$ is a frame over the finite Hilbert space $\mathbb{C}^M$ if its linearly dependent column vectors $\{f_n\}_{n=1}^N \in \mathbb{C}^M$ satisfy:

$$a||y||^2 \leq \sum_{n=1}^{N} |<f_n, y>|^2 \leq b||y||^2, \forall y \in \mathbb{C}^M$$

where the finite coefficients 0<a<b are referred to as frame bounds.

A frame is said to be tight if a=b leaving us with:

$$\sum_{n=1}^{N} |<f_n, y>|^2 = a||y||^2, \forall y \in \mathbb{C}^M$$

Which is equivalent to a frame satisfying:

$$FF^H = aI_M$$

A Unit Norm Tight Frame (UNTF) is a tight frame whose all vectors $\{f_n\}_{n=1}^N \in \mathbb{C}^M$ are unit-norm, which further implicate that $a = b = \frac{N}{M}$ meaning:

$$FF^H = \frac{N}{M} I_M$$

Now we define the vectors' cross correlation:

$$c_{n,k} \stackrel{\text{def}}{=} <f_n, f_k> = f_n' f_k = \sum_{i=1}^{M} F_{i,n}^* F_{i,k}$$

where for UNTF:

$$c_{n,n} = ||f_i||^2 = 1$$

The Welch Bound [3] lower bounds the mean-square cross correlation:

$$I_{ms}(F) \stackrel{\text{def}}{=} \frac{1}{(N-1)N} \sum_{n \neq k} |c_{n,k}|^2 \geq \frac{N-M}{(N-1)M}$$

And is achieved with equality iff F is UNTF.

The Welch Bound implies a bound on the maximum-square error correlation:

$$I_{max}(F) \stackrel{\text{def}}{=} \max_{1 \leq n \neq k \leq N} |c_{n,k}|^2 \geq \frac{N-M}{(N-1)M}$$

We define an Equiangular Tight Frame (ETF) as an UNTF which satisfies:

$$|c_{n,k}|^2 = const = \frac{N-M}{(N-1)M} \quad \forall n \neq k$$

## 3.2. Frames' Singular Values Distribution

Some frames have known asymptotic distribution of their singular values [1]. Those distributions may serve us as a theoretical reference to our results.

### 3.2.1. Random Frames

Let $\tilde{F}$ be a random frame of size $M \times K$ whose entries are random iid variables with zero mean and finite variance. Let's denote a corresponding scaled Gram matrix of this frame by $G = \frac{1}{M}\tilde{F}^H\tilde{F}$ and let $f_\lambda$ be the distribution of eigenvalues of G. Then, at $M, N \to \infty$, $f_\lambda$ converges to Marchenko Pastur distribution:

$$f_\beta(x) = \frac{\sqrt{(\lambda_+ - x)(x - \lambda_-)}}{2\pi\beta x} \mathbb{1}_{[\lambda_- \leq x \leq \lambda_+]}(x)$$

Where $\beta = \frac{K}{M}$ and $\lambda_\pm = \left(1 \pm \sqrt{\beta}\right)^2$

### 3.2.2. Equiangular Tight Frames (ETFs)

Let F be an ETF of size $M \times N$ and let $\tilde{F}$ be a subframe of size $M \times K$ which consists of K randomly chosen columns of F. Let's denote a corresponding scaled Gram matrix by $G = \frac{1}{M}\tilde{F}^H\tilde{F}$ and let $f_\lambda$ be the distribution of eigenvalues of G. Then, at $M, N \to \infty$, $f_\lambda$ converges to Wachter Manova distribution:

$$f_{\beta,\gamma}(x) = \frac{\sqrt{(\lambda_+ - x)(x - \lambda_-)}}{2\pi\beta x(1 - \gamma x)} \mathbb{1}_{[\lambda_- \leq x \leq \lambda_+]}(x) + \left(1 + \frac{1}{\beta} - \frac{1}{\beta\gamma}\right)^+ \delta\left(x - \frac{1}{\gamma}\right)$$

Where $\beta = \frac{K}{M}, \gamma = \frac{M}{N}$ and $\lambda_\pm = \left(\sqrt{\beta(1-\gamma)} \pm \sqrt{1-\beta\gamma}\right)^2$

Above statements are known to hold asymptotically(for large values of M and N). In this paper we'll check if they're still valid for small values of N,M. Moreover they will server us as sort of benchmark when analysing our results.

Here is a graphical visualization of those distributions:

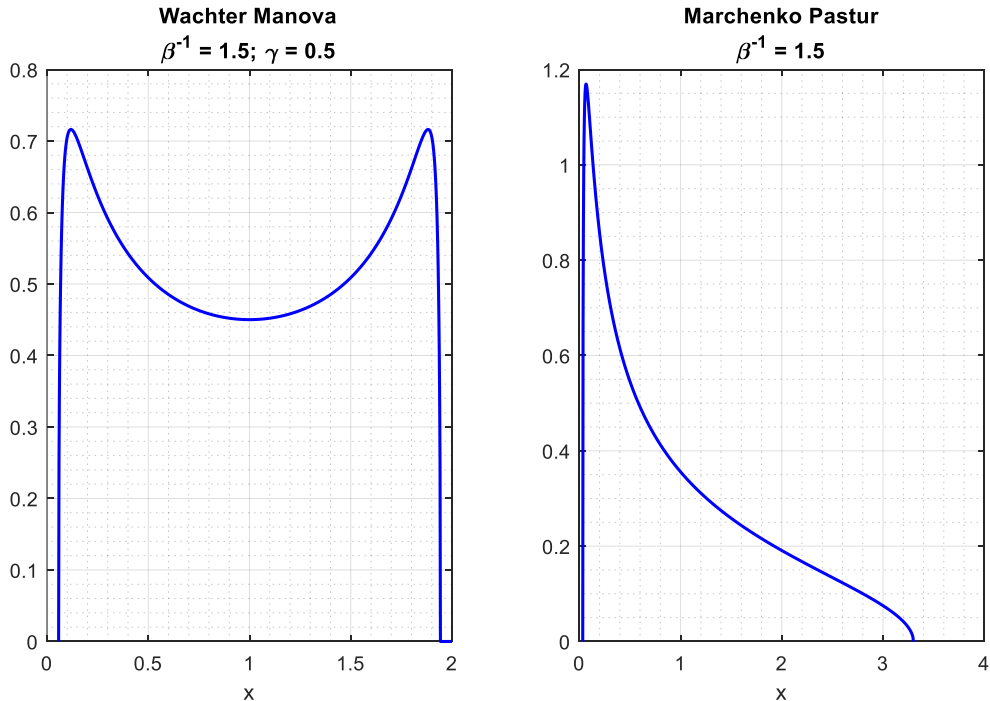

### 3.3. Frames generation

In this paper we'll deal with 2 frames families: random frames and ETF/AETF frames. In both cases we're dealing with binary bipolar frames (+/-1).

Random frames generation is straightforward by producing each element from Bern (0.5) distribution with values 1 or -1.

There exist different approaches for generating ETFs. Among them Steiner ETFs or ETFs generated by taking specific rows of DFT (for complex ETFs) or Hadamard (for binary cases) matrices. A common limitation for all those approaches is that existing ETFs that are created using these methods are limited to some specific values of $\gamma = \frac{M}{N}$.

In [4] it's shown how using difference sets over $(\mathbb{Z}_N, +)$ it's possible to generate complex ETFs by taking specific rows of $N \times N$ DFT matrix. Those rows are chosen according to difference sets.

Using the ideas from [4] and [5] we'll derive similar way of generating binary ETFs by taking specific rows of $N \times N$ Hadamard matrix(which can be thought of as a binary equivalent to DFT matrix). Afterwards we'll extend that idea to generating AETFs.

First let's define what is difference set and what is set's difference spectrum

#### 3.3.1. Difference Set (DS)

A subset D of size M in an abelian group (A,+) with order N is called an (N,M,$\lambda$) difference set (DS) in (A,+) if the multiset $B \stackrel{\text{def}}{=} \{a_1 - a_2 : a_1, a_2 \in A\}$ contains every nonzero element of A exactly $\lambda$ times.

Alternative definition that will be more useful for us is:

A subset D of size M in an abelian group (A,+) with order N is called an (N,M,$\lambda$) difference set (DS) in (A,+) if its difference spectrum(from now on we'll denote it just spectrum) is as follow:

$$\lambda_l = \begin{cases} M, & l = 0 \\ \dfrac{M(M-1)}{N-1}, & l \neq 0 \end{cases}$$

Where we define subset' spectrum as a distribution of its elements differences, namely $\lambda_l$ is defined as follow:

$$\lambda_l = |\{a_1 - a_2 = l : a_1, a_2 \in A\}|$$

From definition of DS spectrum it can be immediately seen that not for any combination of (N,M) there exists difference set, since $\frac{M(M-1)}{N-1}$ must be integer and it is not happening for any (N,M). This constraint limits also ETFs that are generated using DS.

### 3.3.2. Bipolar ETF Generation using Hadamard Matrix

Hadamard matrix is defined as $N \times N$ matrix H with elements as follow:

$$h_{ij} = (-1)^{<i,j>} \quad (3.3.2.1)$$

Where $<i,j> \stackrel{def}{=} \sum_{n=0}^{L-1} b_n^i b_n^j$ is an inner product of 2 binary numbers $b^i$ and $b^j$. $b^i$ is a binary representation of number i and $L = \lceil \log_2 N \rceil$. Note that in standard definition of Hadamard Matrix N is integer power of 2, but for now, we won't constrain ourselves to that condition.

Let's define frame F as a $M \times N$ matrix, which is created by taking M rows from Hadamard Matrix H.

Let's make some denotations:

- $\{u_m\}_{m=1}^M$ is a group of M rows' indices of H that are chosen to create F.
- $f_l$ will denote $l^{th}$ column of frame F

Then the cross correlation between $n^{th}$ and $(n-k)^{th}$ columns of F is given by:

$$c_{n,k} \stackrel{def}{=} f_n^T f_{n-k} = \frac{1}{M} \sum_{m=1}^M (-1)^{<n,u_m>}(-1)^{<n-k,u_m>} = \frac{1}{M} \sum_{m=1}^M (-1)^{<k,u_m>} \quad (3.3.2.2)$$

As we can see $c_{n,k}$ is independent on n so we can just denote it by $c_k$. Note that k's range is $[0, N^+ - 1]$ where $N^+ \stackrel{def}{=} 2^{\lceil \log_2 N \rceil}$ and that is to ensure that $n - k$ covers the whole range of $[0, N-1]$.

Note that only relevant values of $c_k$ are for $k \in \{k | (n-k) \in [0, N-1]\}$.

We have to remember that all arithmetics between indices {n,k,n-l} are over field $GF(2)^L$ – for example subtraction/summation of two indices will be calculated in next way:

$$\{n \pm k\}_{GF(2)^L} = bin2dec(dec2bin(n) \oplus dec2bin(k))$$

Where bin2dec() and dec2bin() are operators that convert binary number to decimal number and vice versa and $\oplus$ represents "XOR" operation.

Let's calculate $c_k^2$:

$$c_k^2 = \frac{1}{M^2} \sum_{i=1}^M (-1)^{<k,u_i>} \sum_{m=1}^M (-1)^{<k,u_m>} = \frac{1}{M^2} \sum_{i=1}^M \sum_{m=1}^M (-1)^{<k,u_i+u_m>} =$$

$$= \frac{1}{M^2} M + \frac{1}{M^2} \sum_{l=1}^{N^+-1} a_l (-1)^{<k,l>} = \frac{N-M}{(N-1)M} + \frac{M-1}{(N-1)M} + \frac{1}{M^2} \sum_{l=1}^{N^+-1} a_l (-1)^{<k,l>} \quad (3.3.2.3)$$

Here, $a_l$ (for $l \in [1, N^+ - 1]$) represents number of times sum $u_i + u_m$ results in $l$ value.

I used the fact that equality $u_i = u_m$ (which is equivalent to $u_i + u_m = 0$) happens exactly M times.

The last sum goes up to $N^+$ and not up to N because $u_i + u_m$ can cause values in range $[0, N^+ - 1]$

Let's define:

$$x_k \stackrel{def}{=} M^2 \left( c_k^2 - \frac{N-M}{(N-1)M} \right) = \frac{M(M-1)}{N-1} + \sum_{l=1}^{N^+-1} a_l (-1)^{<k,l>} = \sum_{l=0}^{N^+-1} a_l (-1)^{<k,l>} \quad (3.3.2.4)$$

Where $a_0 = \frac{M(M-1)}{N-1}$                (3.3.2.5)

We've got that $x_k$ is actually Hadamard Transform of $a_l$.

For frame to be ETF it must hold that:

$$c_k^2 = \begin{cases} 1, & k = 0 \\ \dfrac{N-M}{(N-1)M}, & k \neq 0 \end{cases} \rightarrow x_k = \begin{cases} \dfrac{MN(M-1)}{(N-1)}, & k = 0 \\ 0, & k \neq 0 \end{cases} \quad (3.3.2.6)$$

Namely $x_k$ is a delta function – that means that it's inverse Hadamard Transform has to be constant, namely $a_l$ has to be constant. Given (3.3.2.5) we get

$$a_l = \frac{M(M-1)}{N-1} \; \forall l \quad (3.3.2.7)$$

We remind that $a_l$ (for $l \in [1, N^+ - 1]$) represents number of times sum $u_i + u_m$ results in $l$ value – namely, number of time difference l is present in a given set $\{u_i\}_{i=1}^M$, and together with (3.3.2.7) it leads us to conclusion that $\{u_i\}_{i=1}^M$ is a difference set.

Let's validate that inverse Hadamard Transform of $x_k$ is indeed equal to (3.3.2.7):

$$\frac{1}{N^+} \sum_{k=0}^{N^+ - 1} x_k (-1)^{<k,l>} = \frac{1}{N^+} \frac{MN(M-1)}{(N-1)} \; \forall l \quad (3.3.2.8)$$

We see that (3.3.2.7)= (3.3.2.8) iff $N^+ = N$ which is the case when N is an integer power of 2. In that case we get that in order to get ETF from Hadamard Matrix (with N being integer power of 2) we have to take H's rows with indices that form difference set (over $GF(2^L)$).

### 3.4. Basics of Genetic Algorithms (GA)

In section 4.1.1. we'll make use of Genetic Algorithms(GA) to generate Generalized Difference Sets(GDS)(which will be defined in advance). Hence we'll give here a brief presentation of basic steps of GA.

Genetic Algorithm (GA) is an optimization technique inspired by the process of a natural selection.

The main steps of GA are:

1. Initial random population generation. Each individual in this population is usually represented as a binary string. Each individual is assigned some fitness value (which is obtained by fitness function)
2. Selection: the fittest individuals (individuals with best fitness values) are chosen for creating next generation of individuals.
3. Crossover: process of merging between individuals (between their binary strings) to produce a new ones for a next generation.
4. Mutation: process of randomly altering some individuals' bits. The main goal of that step is to present randomness to the whole process.
5. Calculate fitness for each of the new individuals and go back to Selection Step. The whole procedure will continue until predefined number of iterations is achieved.

# 4. A Presented Method: Bipolar Almost ETFs Generation using Hadamard Matrix

In section 3.3.2. we've seen that Hadamard Matrix may be used to generate ETFs only when N is an integer power of 2.

What happens when N is not power of 2? We still can use above technique with small modifications to generate frames that will have properties which are similar to ETFs. We'll call those frames Almost ETFs or AETFs.

More precisely, we'll change our requirement on $c_k^2$ and instead of requiring pure ETF (3.3.2.6), we'll require:

$$c_k^2 = \begin{cases} 1, & k = 0 \\ \dfrac{N-M}{(N-1)M}, & k \in [1, N^- - 1] \\ \dfrac{N-M}{(N-1)M} + \alpha, & k \in [N^-, N^+ - 1] \end{cases} \rightarrow x_k = \begin{cases} \dfrac{NM(M-1)}{N-1}, & k = 0 \\ 0, & k \in [1, N^- - 1] \\ \alpha M^2, & k \in [N^-, N^+ - 1] \end{cases} \quad (4.1)$$

Where $N^- = \frac{1}{2}N^+ = \lfloor \log_2 N \rfloor$ and $\alpha$ is some positive constant. In our analysis we're dealing with range: $N^- < N \leq N^+$

This specific division to k's regions is not arbitrary and comes from the next observation:

- For $N^- \leq k < N^+$ $c_k$ is corresponding to crosscorrelations between first $N^-$ and last $N - N^-$ columns of F – that means in total $N^- \leq 2N^-(N - N^-) \leq 2(N^-)^2$ cross correlations.
- For $1 \leq k < N^-$ $c_k$ is corresponding to the rest possible cross correlations of F's columns – that means in total $N^-(N^- - 1) \leq N(N-1) - 2N^-(N - N^-) \leq 2N^-(N^- - 1)$ cross correlations.

In both cases upper bound is achieved when $N = N^+$. We can see that only at upper bound $c_k$ appearance is almost uniformly for all k's. For $N < N^+$ $c_k$ is much more frequent for $k \in [1, N^- - 1]$ then for $k \in [N^-, N^+ - 1]$. That's the reason we've chosen $c_k^2$ to be smaller (and closer to ideal ETF) for $k \in [1, N^- - 1]$.

The next step to formulize the definition is to understand what value should $\alpha$ take.

Let's calculate $a_l$ corresponding to given $x_k$ (4.1):

$$a_l = \frac{1}{N^+} \sum_{k=0}^{N^+ - 1} x_k (-1)^{<k,l>} = \frac{1}{N^+} \sum_{k=0}^{N^- - 1} x_k (-1)^{<k,l>} + \frac{1}{N^+} \sum_{k=N^-+1}^{N^+ - 1} x_k (-1)^{<k,l>} \quad (4.2)$$

We'll divide calculation to 2 regions:

$$a_{l \in [0, N^- - 1]} = \frac{1}{N^+} \left[ \frac{NM(M-1)}{N-1} \right] + \frac{1}{N^+} \left[ \sum_{k=0}^{N^- - 1} \alpha M^2 (-1)^{<k,l>} \right] \quad (4.2a)$$

$$a_{l \in [N^-, N^+ - 1]} = \frac{1}{N^+} \left[ \frac{NM(M-1)}{N-1} \right] - \frac{1}{N^+} \left[ \sum_{k=0}^{N^- - 1} \alpha M^2 (-1)^{<k,l>} \right] \quad (4.2b)$$

We'll require (as in (3.3.2.5)):

$$a_0 = \frac{M(M-1)}{N-1} \quad (4.3)$$

On the other hand we've got (4.2a):

$$a_0 = \frac{1}{N^+}\left[\frac{NM(M-1)}{N-1}\right] + \frac{N^-}{N^+}\alpha M^2 \quad (4.4)$$

From comparing last two results we get:

$$\alpha = 2\frac{(M-1)\left(1 - \frac{N}{N^+}\right)}{M(N-1)} \quad (4.5)$$

It can be bounded by inserting $N = N^-$ for upper bound and $N = N^+$ for lower bound:

$$0 \le \alpha = 2\frac{(M-1)\left(1 - \frac{N}{N^+}\right)}{M(N-1)} < \frac{(M-1)}{M(N-1)} \quad (4.6)$$

Inserting $\alpha$ that we've got in (4.5) into (4.1) gives us crosscorrelation:

$$c_k^2 = \begin{cases} 1, & k = 0 \\ \frac{N-M}{(N-1)M}, & k \in [1, N^- - 1] \\ \frac{N-M}{(N-1)M} + 2\frac{(M-1)\left(1 - \frac{N}{N^+}\right)}{M(N-1)}, & k \in [N^-, N^+ - 1] \end{cases} \quad (4.7)$$

Using above bounds (4.6) the last term of $c_k^2$ becomes $\frac{N-M}{(N-1)M}$ for $N = N^+$ bringing us back to pure ETF case, and in worst case it is upperbounded by:

$$\frac{N-M}{(N-1)M} + \frac{(M-1)}{M(N-1)} = \frac{1}{M} \quad (4.8)$$

By inserting resulting $\alpha$ into $a_l$ ($l \in [1, N^+ - 1]$) (4.2) we get next desired $\lambda$'s spectrum of the set $\{u_m\}_{m=1}^M$:

$$\lambda_l = \begin{cases} M, & l = 0 \\ \left(2\frac{N}{N^+} - 1\right)\frac{M(M-1)}{N-1}, & l = N^- \\ \frac{N}{N^+}\frac{M(M-1)}{N-1} & otherwise \end{cases} \quad (4.9)$$

Which is again becomes spectrum of pure difference set for $N^+ = N$.

Not that for general N it is still very close to spectrum of difference set for $l \ne N^-$. We'll denote set with spectrum (4.9) as a Generalized Difference Set (GDS).

## 4.1. Generalized Difference Sets (GDS) generation

Till now we have that problem of construction of AETFs is reduced to a problem of construction of GDS with differences spectrum given by (4.9)

Construction of classical difference sets is a complex maths problem and a lot of papers were written on this topic proposing various techniques, when each of them allows to construct only specific subfamily of difference sets.

In our case we're dealing with set which is not classical difference set and which has very specific difference spectrum. To our knowledge, there is no technique for construction of that type of set.

Inspired by ideas of [6] we suggest to use Genetic Algorithm for construction of this kind of sets. The big advantage of Genetic Algorithm is its generic nature which is "blind" to details of specific problem and it is only concerned about optimizing "Loss Function". That allows us to use it to look for desired sets. The main disadvantage of Genetic Algorithm is that convergence to a global optimum is not guaranteed – moreover due to its probabilistic nature convergence may be different in two different runs of algorithm. Despite of this drawback Genetic Algorithm still has a potential to give us a good results.

Note also that even if Genetic Algorithm does converge to a global optimum – it doesn't mean we'll get desired spectrum, because for some $(M, N, N^+)$ desired spectrum may be unachievable (for example if some of $\lambda_l$ values is not an integer) – in those case the best we can get is an approximation to desired spectrum.

### 4.1.1. Applying GA to a problem of searching new GDS

We'll go through the steps described in previous subsection and describe how those are implemented in our problem.

1. We start with N randomly generated sets. Each set is an individual and all N generated sets are the whole population. A binary representation of every set is a $N^+$-length binary string with ones at indices, which are members of a given set. For example if $N^+ = 8$, M = 3 and the set is {1,4,6}, the binary representation of that set will be (1,0,0,1,0,1,0,0).

   Fitness Function Definition

   Our final goal is to get set with spectrum given by (4.9). Let's denote that desired spectrum by $\underline{\lambda}^*$. If $\underline{\lambda}$ is a spectrum of a given set than fitness function is defined as:

   $$F(\underline{\lambda}) = \alpha|\lambda(N^-) - \lambda^*(N^-)|^2 + \beta\left\|\underline{\lambda} - \underline{\lambda}^*\right\|^2$$

   Where $\alpha = 1$ and $\beta = 10^{-4}$ – we use those coefficients to give higher weight to optimization of $\lambda_l$ at $l = N^-$.

   We calculate that function for each of the individuals.

2. Selection: $\frac{N}{2}$ pairs that are supposed to be parents for next generation are chosen randomly where probability of choosing of specific individual is inversely proporcional to its fitness value. Same individual may be chosen a number of times as a part of a different pairs.
3. Crossover: A common crossover technique in GAs is given a pair of individuals(parents), randomly choose index of their binary representations and create two new (children) individuals by merging their binary representation.
   For example, given two parents:
   $$p_1 = \{0,1,1,0,0,0\}; p_2 = \{0,0,1,0,1,0\}$$
   And given that index that was chosen is 2 the children will be:
   $$c_1 = \{0,1,1,0,1,0\}; c_2 = \{0,0,1,0,0,0\}$$
   In our case though this crossover technique may be problematic as it can be seen from that simple example, number of one's in a binary representation may change from generation to generation, namely M (-number of sets' elements) will change from generation to generation. This is not good, because we're interested in a predefined set size.

Hence we need a crossover technique that will preserve the number of "ones" bits in a representation. The way we do it is by merging set of "ones" indices of both parents, permute them randomly and choosing first M indices of permuted set as "ones" indices of a first child and last M indices of permuted set as "ones" of a second child.

For example given parents as above, merged set of "ones" indices of both parents is:
$$\{2,3,5\}$$
After random permutation we can get for example {3,2,5}. In that case 1st child will get first 2 indices and 2nd child will get last 2 indices:
$$c_1 = \{0,1,1,0,0,0\}; c_2 = \{0,1,0,0,1,0\}$$
Note that this technique ensures that number of "ones" bits is preserved between generations.

One more thing that we do is calculating fitness function of each of the children, and then among set {parent1,parent2,child1,child2} we choose for next generation a pair with a best fitness. In such a way we ensure that fitness of a next generation is at least as good as a fitness of a previous generation.

Crossover step happens with probability 0.9.

4. Mutation: Common mutation approach is randomly flipping single bit of a binary sequence, but in our case again to preserve the constant number of ones, we'll always flip bits in pairs – for every flipping "1->0" we perform also flipping "0->1".
Mutation step happens with probability 0.1.
5. Calculate fitness of a new generation and go back to step 2.

# 5. Performance Metrics

We'll use the same metrics as in [2]

1. Capacity – the channel capacity of a linear Gaussian Channel is defined to be:

$$C\left[\frac{bits}{s}\right] = E\{\log_2(\det(I + SNR \cdot \tilde{F}^H \tilde{F}))\} = E\left\{\log_2 \prod_{i=1}^{K}(1 + SNR \cdot \lambda_i)\right\}$$

$$= E\left\{\sum_{i=1}^{K} \log_2(1 + SNR \cdot \lambda_i)\right\} \quad (4.1)$$

Where $\lambda_i$ are the eigenvalues of $\tilde{F}^H \tilde{F}$.

2. Practical Capacity. Above mentioned capacity is a theoretical limit which can be achieved but it requires complex implementations – hence it's less relevant for most practical systems. Instead it's more convenient to presents slightly different measure which is more aligned with performance of a practical relatively simple systems. We'll refer it as a practical capacity:

$$C_p\left[\frac{bits}{s}\right] = E\{\log_2(\det(SNR \cdot \tilde{F}^H \tilde{F}))\} = E\left\{\sum_{i=1}^{K} \log_2(SNR \cdot \lambda_i)\right\} \quad (4.2)$$

Each of the above metrics can be normalized per user $\frac{C}{K}$ or per resource $\frac{C}{M}$. Those normalizations will be useful for us in our analysis.

# 6. Results and Analysis

As we've already mentioned the frames that are generated used technique presented in this paper are not guaranteed to be optimal in sense of capacity because:

- The main approach is dealing with AETFs instead of ETFs(besides very specific cases where N is a power of 2) – hence it's not guaranteed to achieve Wachter-Manova Performance which is performance of a pure ETFs.
- As was mentioned above Genetic Algorithm is not guaranteed to converge to global optimum and when it does, it still doesn't guarantee that we'll get desired AETF.

Hence our analysis is concentrating on:

- Quantifying the misoptimality of our frames by comparing their performance to theoretical performance of ETFs derived from Wachter Manova distribution.
- Showing if our frames are still present some performance benefits in comparison to binary random iid frames.

We'll present capacity per user performance versus different $p, \beta^{-1}$ and N values. We'll look both at theoretical capacity and at practical one.

<u>Simulations Settings</u>

- Simulation was splitted in 2 steps:
    - GDS generation for different values of N and M using Genetic Algorithm
    - Capacity simulation of iid frames and AETFs that were generated using GDS from the first step.
- AETF and iid frames capacities are calculated according to definitions in section 4.
- Marchenko Pastur and Manova capacities are calculated using theoretical distributions presented in section 3.2.
- All capacities calculations of AETF frames are averaged on 1000 iterations of randomly taking K out of N columns of a given AETF. Note that for any given K we're taking exactly K F's columns(as opposite to alternative option where we could take each column with probability $\frac{K}{N}$)

## Capacity Per User Versus N

$$\beta^{-1} = 1.25$$

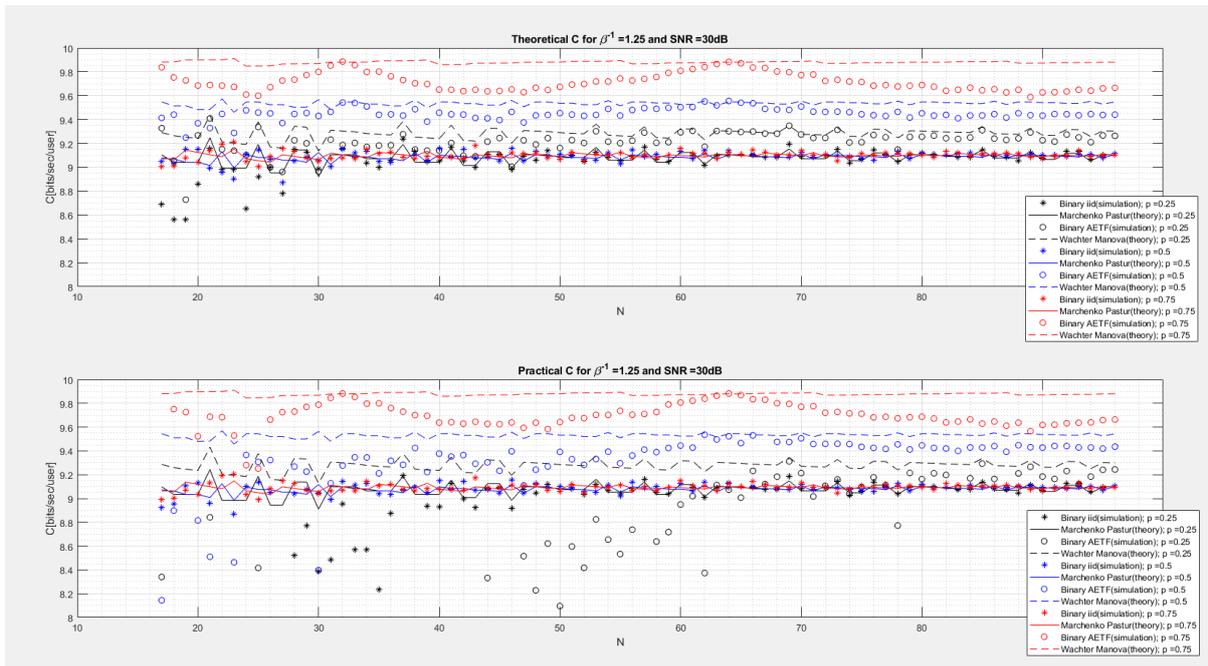

$$\beta^{-1} = 1.5$$

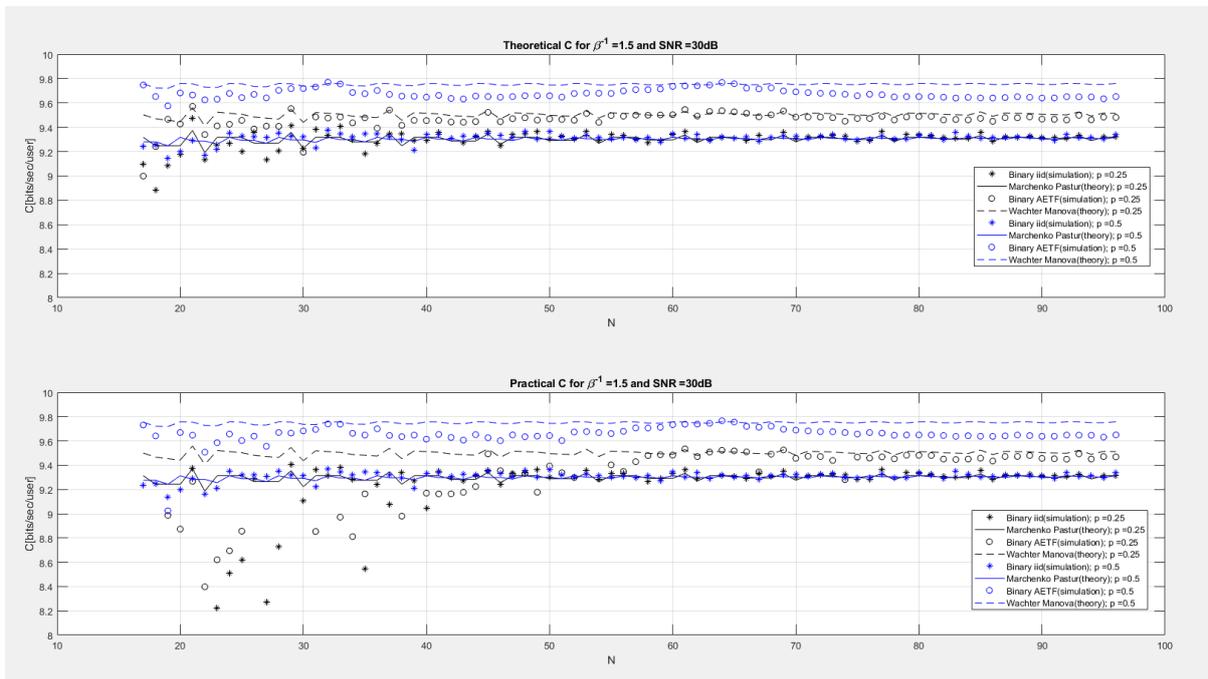

$$\beta^{-1} = 1.75$$

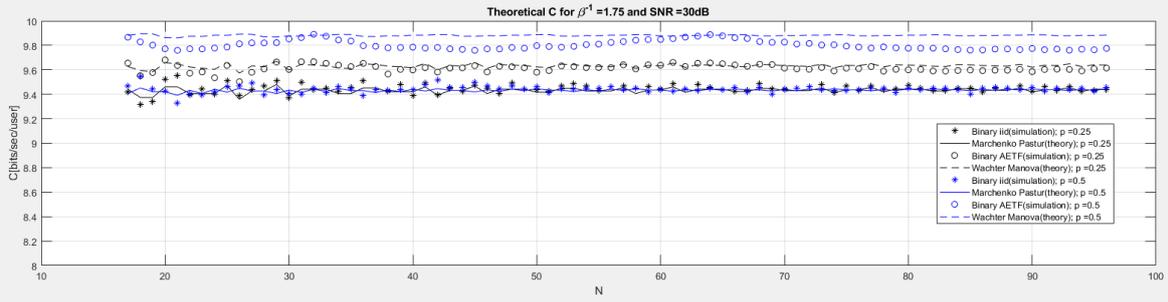

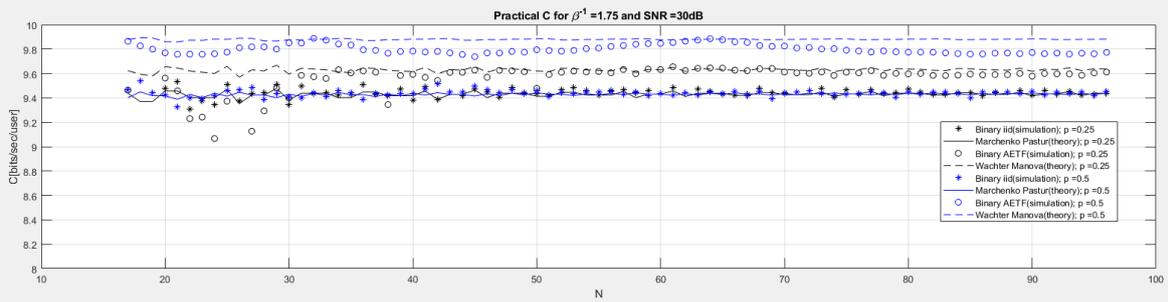

# Capacity Per User Versus $\beta^{-1}$ for some Ns

## N=17

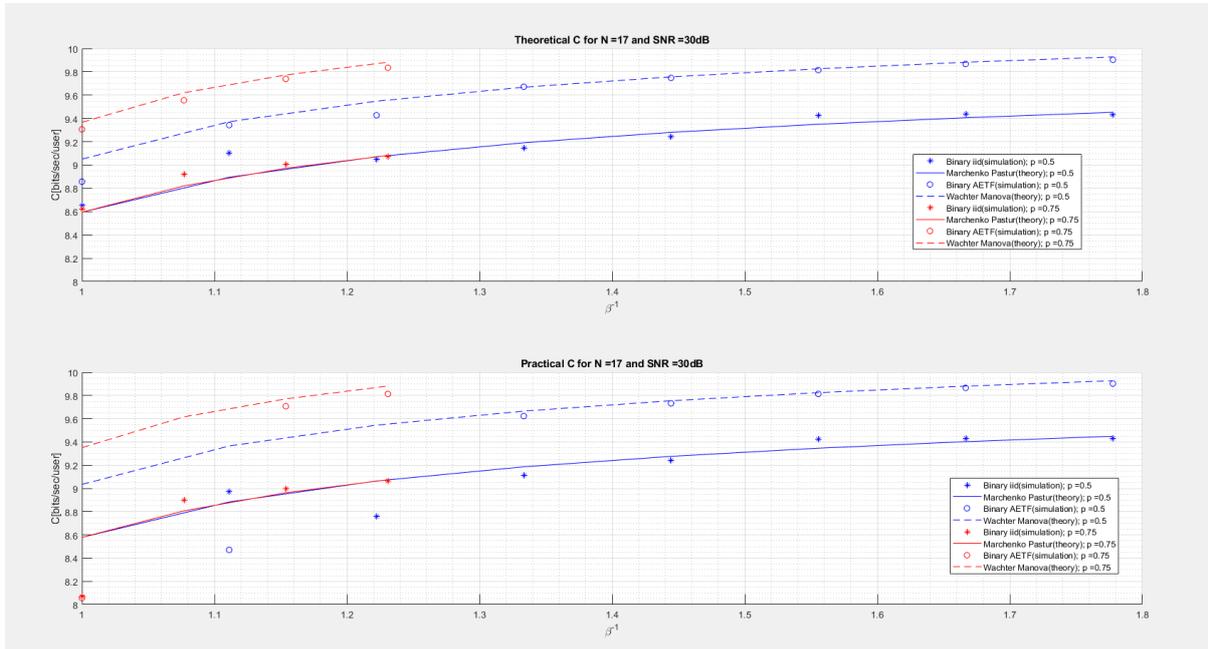

## $N = 32$

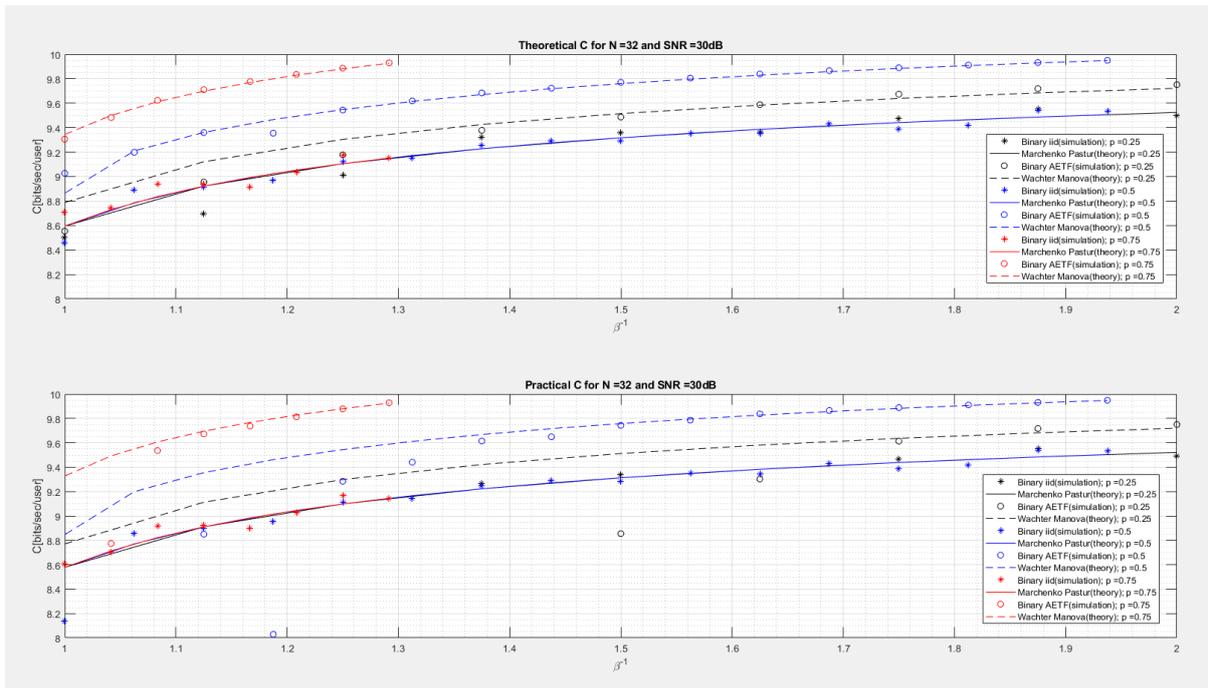

N=48

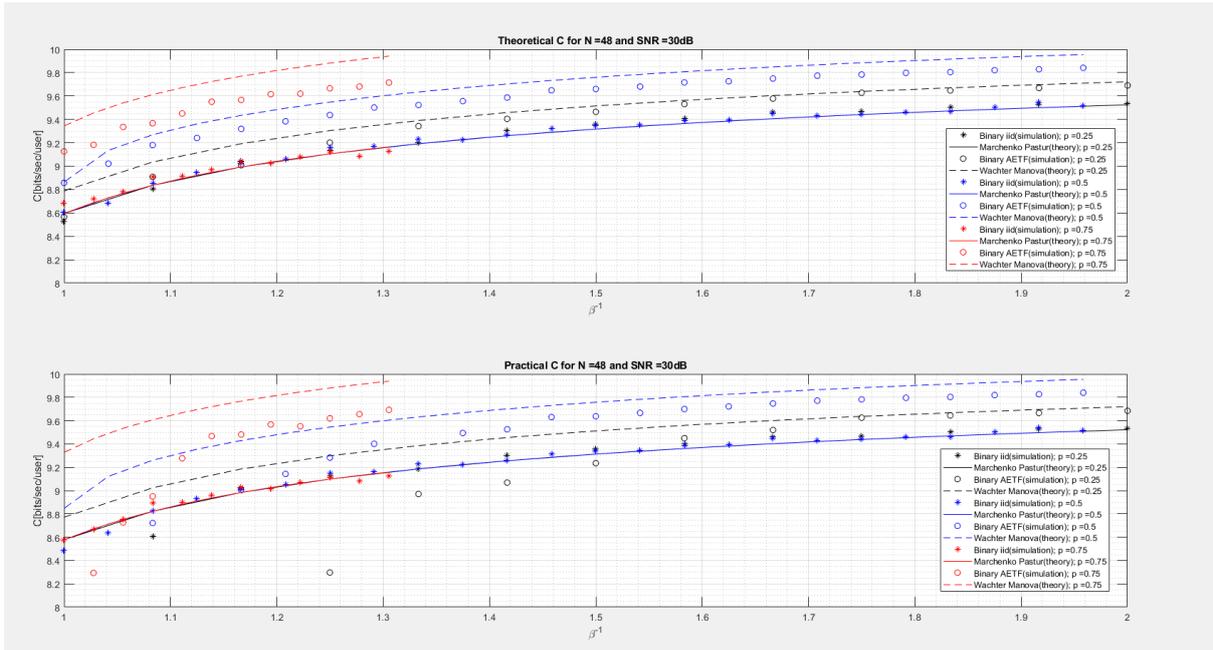

N=96

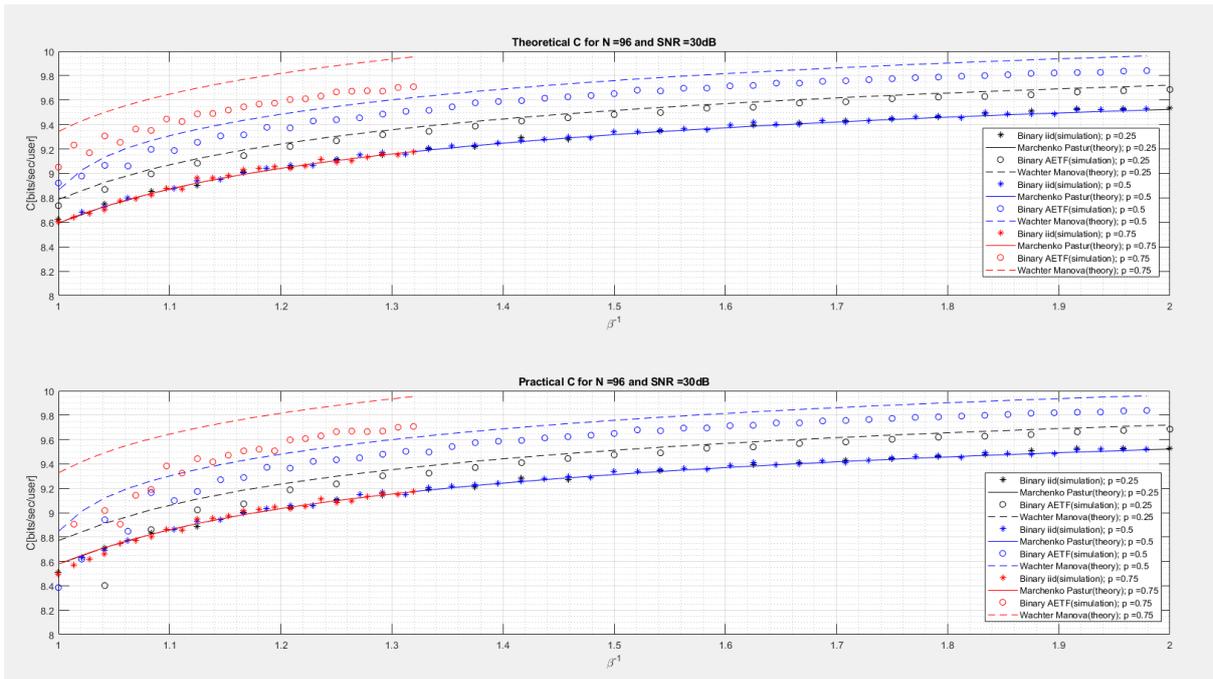

Results Analysis

Let's discuss theoretical capacity first:

- First of all, we can see that as expected we have the best performance for N's that are integer power of 2 and as we move away from it the performance gets worse. This is due to the fact that we don't get ETFs but only AETFs for those values. Note also, that for N that are power of 2 the performance of our frames is coincides with Wachter Manova performance.
- With that being said, we can observe that in all scenarios and for all Ns, ETF/AETF frames outperform random binary frames. Moreover, it's capacity loses to Wachter Manova by at most 0.3[bits/sec/user].
- One more interesting observation is that binary iid frames' performance almost coincides with the Marchenko-Pastur distribution for N>20, which is not so obvious, because the Marchenko Pastur distribution was initially derived as an asymptotic distribution.

Practical Capacity:

- Regarding practical capacity, AETF is outperform random iid frames only in specific scenarios. Next table summarizes in what cases it'll be beneficial to use AETFs and not binary iid frames:

|  | p=0.25 | p=0.5 | p=0.75 |
|---|---|---|---|
| $\beta^{-1} = 1.25$ | $N > 80$ | $N > 30$ | *all N*$^*$ |
| $\beta^{-1} = 1.5$ | $N > 55$ | *all N*$^*$ | -- |
| $\beta^{-1} = 1.75$ | $N > 30$ | *all N*$^*$ | -- |

*all N – meaning "all tested Ns". Range from N=16 to 96.

- In most cases where iid frames outperform AETF frames we can see a significant drop in performance of practical capacity in comparison to theoretical capacity. The reason to such a degradation in performance is that for practical capacity zero eigenvalues causes in explosion of log function as can be seen from (4.2).

# 7. Conclusions and Future Work

In that paper we've presented a way of generating binary AETFs using Hadamard Matrix and GDS. This way allowed us to generate frames for arbitrary values of N and M. The performance of those frames was analyzed and it was shown that if we're looking at theoretical capacity, those frames are outperform binary iid random frames, moreover, their performance is not far from real ETF performance. If we're looking at practical capacity, AETFs are still can be beneficial but only for specific values of $\beta^{-1}, p, N$ which were summarized.

Possible directions for improvement of the results:

- $c_k^2$ that we've derived in (3.3.3.7) is constrained to only 2 values which is not necesserely the optimal constraint.
- Genetic Algorithm as it was pointed out is not guaranteed to converge to global optimum. Hence improving ways of looking for GDS may present some additional improvement in performance.
- We're dealed with a very specific way of generating ETFs (using Hadamard Matrices and Difference Sets). Other ways may have some benefits, for example ability to generate binary ETFs for Ns that are not integer power of 2.

In additional to improvement of the results, it may be interesting in future to investigate unipolar frames (0,1 valued frames) and maybe even trinary frames (1,0,-1) as those types frames may allow even simpler implementation due to presence of zeros.

# 8. References

[1] Marina Haikin, Matan Gavish, Dustin G. Mixon and Ram Zamir, "Asymptotic Frame Theory for Analog Coding," 2021.

[2] Maya Slamovich, Ram Zamir, "Frame-based codes for partially active NOMA," 2021.

[3] Welch, L., "Lower bounds on the maximum cross correlation of signals," 1974.

[4] Pengfei Xia, Shengli Zhou, G.B. Giannakis, "Achieving the Welch bound with difference sets," 2005.

[5] Ding, Cunsheng, "A Generic Construction of Complex Codebooks Meeting the Welch Bound," 2007.

[6] G. Oliveri, M. Donelli, A. Massa, "Genetically-designed arbitrary length almost difference sets," 2011.